\documentclass[12pt,a4paper]{article}
\usepackage[cp866]{inputenc}
\usepackage{amsmath}
\usepackage{graphicx}
\usepackage{epsfig}
\usepackage{subfigure}
\newcommand{\be}{\begin{equation}}
\newcommand{\ee}{\end{equation}}
\newcommand{\ben}{\begin{equation*}}
\newcommand{\een}{\end{equation*}}
\newcommand{\bea}{\begin{eqnarray}}
\newcommand{\eea}{\end{eqnarray}}
\newcommand{\ar}{\begin{array}}
\newcommand{\arn}{\end{array}}

\newcommand{\vk}{\vec{k}}
\newcommand{\vks}{\vec{k}^{\;2}}
\newcommand{\q}{\vec{q}}
\newcommand{\qs}{\vec{q}^{\;2}}

\newcommand{\bor}{\vec{r}}
\newcommand{\boq}{\vec{q}}
\newcommand{\bol}{\vec{l}}
\newcommand{\bok}{\vec{k}}

\def\pnot{\mbox{${\not{\hbox{\kern-3.0pt$p$}}}$}}
\def\qnot{\mbox{${\not{\hbox{\kern-2.0pt$q$}}}$}}
\def\enot{\mbox{${\not{\hbox{\kern-2.0pt$e$}}}$}}
\def\knot{\mbox{${\not{\hbox{\kern-2.0pt$k$}}}$}}

\def\fun#1#2{\lower3.6pt\vbox{\baselineskip0pt\lineskip.9pt\ialign
{$\mathsurround=0pt#1\hfil##\hfil$\crcr#2\crcr\sim\crcr}}}

\begin{document}
\sloppy
\renewcommand{\baselinestretch}{1.0} 
	
\begin{titlepage}
		
\begin{center}
{\bf BFKL -- past and future$^{\ast}$}
\end{center}
\centerline{V.S.~Fadin$^{a, b\,\dag}$}

\vskip .6cm

\centerline{\sl $^{a}$Budker Institute of Nuclear Physics of SB  RAS, 630090 Novosibirsk, Russia}
\centerline{\sl $^{b}$ Novosibirsk State University, 630090 Novosibirsk, Russia}

\vskip 2cm		

\begin{abstract}
This paper contains my recollection  of the creation and development of the so-called BFKL approach and my ideas about the ways of its further development.
\end{abstract}


\vfill \hrule \vskip.3cm \noindent $^{\ast}${\it Work supported
	in part by the Ministry of  Science and Higher Education of Russian Federation and
	in part by  RFBR,   grant  19-02-00690.} \vfill $
\begin{array}{ll} ^{\dag}\mbox{{\it e-mail address:}} &
\mbox{fadin@inp.nsk.su}\\
\end{array}
$
\end{titlepage}

\vfill \eject

\section{BFKL history}
The BFKL story  goes back to the middle of 70-th of the last century, when Lev  started \cite{Lipatov:1976zz} to study the high energy behaviour of non-Abelian gauge theories \cite{Yang:1954ek}. To avoid infrared singularities,  the investigation  was performed in the theories with    Higgs mechanism of  mass generation \cite{Higgs:1966ev} - \cite{Englert:1964et} preserving renormalizability.  More definitely, the model  conserving global isotopic invariance was considered, with local gauge symmetry SU(2) and isotopic doublet of scalar fields. The paper \cite{Lipatov:1976zz} was devoted to check  of Reggeization of the gauge boson  and investigation of the vacuum singularity in this theory.


\subsection{Historical background}
Recall that the notion of Reggeization of elementary particles in perturbation theory was introduced and study of this phenomenon  was started in the series of papers by Gell-Mann and co-authors \cite{Gell-Mann:1962xxa} - \cite{GellMann:1964zz}. In terms of the relativistic partial wave amplitude $A_j(t)$, analytically continued to complex $j$
values, the Reggeization was defined as the disappearance, due to
radiative corrections,  of  the non-analytic terms of  the Born approximation on account
of  one-particle exchanges in the $t$ channel. In other words,  Reggeization
of an elementary particle with  spin $j_0$ and  mass $m$ means that
at large $s$ and fixed  $t$ Born  amplitudes with exchange of this
particle in the $t$ channel  acquire a factor $s^{j(t)-j_0}$,
with  $j(m^2)=j_0$, as a result of  radiative corrections.  Gell-Mann and his collaborators established the presence of this phenomenon  for backward Compton scattering  in QED with heavy photon   and thus showed the fermion Reggeization in this theory. They also investigated scalar electrodynamics and concluded that a scalar is not Reggeized.  In \cite{Gell-Mann:1964aya} they  formulated necessary for Reggeization  conditions  on  the Born scattering amplitudes of the theory. These conditions were generalized later in \cite{Mandelstam:1965zz}. In \cite{Mandelstam:1964tvk} it was shown that they are not fulfilled in the vector  channel  of massive QED. Non-Reggeization of the heavy photon was confirmed by direct two-loop calculations in \cite{Chester:1965ey}.

Note that the term "Reggeization" can be understood in different ways.
If we assume that it means only the existence of the Regge trajectory on which the particle lies, then the scalar interacting with the photon is Reggeized \cite{Cheng:1974sv}. We use this term in a stronger sense: it means not only the existence of the Regge trajectory of the particle, but that the entire amplitude  is given by the Reggeon with  this trajectory.

In non-Abelian gauge theories the Reggeization problem  was considered in Refs. \cite{Abers:1970wn} - \cite{Grisaru:1976ey}.  In particular, it was shown in  \cite{Grisaru:1973vw}, \cite{Grisaru:1974cf}  that in the theories  with the  Higgs mechanism of mass generation (unlike the theories with mass terms in original Lagrangian \cite{Dicus:1972rt}) the necessary for Reggeization conditions  \cite{Gell-Mann:1964aya} - \cite{Mandelstam:1965zz} are fulfilled for  the  vector  channel.   Fulfilment of these conditions allowed  to find  \cite{Grisaru:1973vw}  the trajectory of the vector meson  using the $t$-channel unitarity:
\be
j_V(t)= 1+ \frac{8g^2}{\pi}(t-m_V^2)\int_{4m_V^2}^\infty \frac{dt'}{t'-t}\frac{\rho(t')}{t'-4m_V^2}~, \label{trajectory} 
\ee
where
\be
\rho(t) =\frac{1}{32\pi}\sqrt{\frac{t-4m_V^2}{t}} ~. \label{jump}
\ee
However, this was  only an indication of the possibility of the Reggeization, and not at all a proof of it.  Therefore, it was very important to check the Reggeization by direct calculations, and  the first Lev's aim   was to do it.

The second aim was to investigate  the vacuum singularity, i.e. asymptotics of amplitudes   with vacuum quantum numbers in the $t$-channel and positive signature.
In the Regge-Gribov theory of complex angular  momenta this singularity  originally  was introduced (with intercept equal to one) \cite{Gribov:1961fr}, \cite{Chew:1961ev}
to provide constant cross sections at asymptotically high energies. Because of its fundamental role, this singularity  has received a special name:
it was  called Pomeron after I.\,Ya.~Pomeranchuk.

This singularity  was investigated in QED \cite{Gribov:1970ik} -  \cite{Cheng:1970xm}
 (Lev  was one of the main investigators),  and it was shown that  in the LLA it is a fixed branch point to the right of 1 
(at $j= 1+\frac{11}{32}\pi\alpha^2$),  i.e. the LLA violates the Froissart bound \cite{Froissart:1961ux}. Investigation of the Pomeron   in the non-Abelian gauge theories was especially interesting because of vector boson Reggeization, unlike photon.

\subsection{Dispersive approach and the first results}
It was important to check the Reggeization by direct calculations, and it was done  in  \cite{Lipatov:1976zz} in two loops with the leading logarithmic accuracy (LLA), when in each order of perturbation theory only terms with the highest powers of $\ln s$ are kept.

For the calculations  Lev used  the dispersive approach based on the general properties of the theory: analyticity, unitarity and renormalizability. To my mind, it was the first application of the dispersive approach to the non-Abelian gauge theories. This approach turned out very successful, because it permits to escape consideration of great number of Feynman diagrams and to work only with physical particles in the unitarity relations thus avoiding the use of ghosts by Faddeev-Popov. Now it is widely used, unfortunately, without any reference to Lev.

Note that the term LLA was used  to indicate that only terms with the highest powers of the logarithm of c.m.s.  energy $\sqrt s$ are  being held  in amplitude discontinuities, not in the amplitudes themselves. Total  amplitudes are obtained from their $s$-channel discontinuities in this approximation by  the substitutions
\be
is\ln^n s  \rightarrow -\frac{s}{2\pi (n+1)} \left(\ln^{n+1}(-s-i\epsilon)\pm  \ln^n (s-i\epsilon)\right)~, \label{connection}
\ee
where $+(-)$ sign for negative (positive) signature (symmetry with respect to the replacement $s\leftrightarrow u\simeq  - s$). Therefore  the degrees  of logarithms in the real parts of the amplitudes with a negative signature are  one more than the degree in the imaginary parts, whereas for the amplitudes with a positive  signature, the reverse is true, and in the LLA they are purely imaginary. It is very important, because thanks to this in the LLA (and in the next to it, as it will be discussed below)  only amplitudes with negative signature must be kept  on the right side of the unitarity relations
\be
\mbox{Im}_{s}{\cal A}^{A'B'}_{AB}\ =\ \frac12 \sum_{n=0}^\infty \sum_{\{f\}} \int {\cal
A}^{\tilde A  \tilde B +n}_{AB} \left({\cal A}^{\tilde A  \tilde B +n}_{A'B'}\right)^*
d\Phi _{\tilde A \tilde B +n}\,, \label{elastic discontinuity}
\ee
where $\sum_{\{f\}}$ means sum over discrete quantum numbers of intermediate
particles, ${\cal A}^{\tilde A
 \tilde B +n}_{AB}$ and
$\left({\cal A}^{\tilde A  \tilde B +n}_{A'B'}\right)$ are the amplitudes of $n$-particle  production and $d\Phi_{\tilde A \tilde B +n}$ is the corresponding phase
space element.

As it is seen from (\ref{elastic discontinuity}), calculation of  elastic amplitudes in the dispersive approach requires knowledge  of inelastic  ones. It was recognized by Lev that the inelastic amplitudes are necessary in special kinematics. In the LLA it is  so called multi-Regge kinematics (MRK), where produced particles have limited (not growing with $s$) transverse momenta and are strongly ordered in rapidity.  For two-loop  calculations one-particle production amplitudes in the Born approximation  are  necessary. Using the $t$-channel unitarity, Lev obtained a  simple factorized form of these amplitudes with famous now Lipatov's vertex for production of vector mesons.

The main result of Ref.~\cite{Lipatov:1976zz} was  that in the LLA two-loop radiative corrections to elastic scattering amplitudes with the gauge boson quantum numbers  in the $t$-channel and  negative signature  have the Regge  form. As for the vacuum singularity, it was shown only that it has not a pole form. More definite conclusions about the nature of this singularity on the basis of two-loop  calculations can not be drawn.

\subsection{Pre-BFKL}

Next important steps in the investigation of the  high energy behaviour of non-Abelian gauge theories was done in \cite{Fadin:1975cb} - \cite{Kuraev:1977fs}. Two-particle production amplitudes were found in the Born approximation and the one-loop corrections to the one-particle production amplitudes were calculated. It turned out that the  first ones  have the factorized form with the same vertices as the  one-particle production amplitudes and that the corrections to the last ones  have the Regge  form. Three-loop corrections to elastic amplitudes  calculated using these results proved the vector boson Reggeization in this order. Details of  derivation of all these results and their generalization on colours group $SU(N)$ with arbitrary $N$ were  described  in \cite{Kuraev:1976ge}. 

An extremely important step made on this basis was the hypothesis of vector meson Reggeization.    It was assumed that in  the LLA not only elastic, but also  inelastic amplitudes in the MRK with quantum numbers of vector bosons and negative signatures in all cross-channels  are given by  the  Regge pole  contributions in all orders of perturbation theory. Since only such amplitudes are important in the unitarity relations (\ref{elastic discontinuity}), it is possible to express  partial waves $(A_{AB}^{A'B'})^T_\omega (t)\;\;\;  (\omega = j-1)$ of amplitudes of the process  ${AB}\rightarrow {A'B'}$ with the global $SU(2)$ "isospin" $T$  
\[
(A_{AB}^{A'B'})^T_\omega (t) = \int_{s_{0}}^\infty\frac{ds}{s} s^{-\omega -1}  \mbox{Im}_{s} (A_{AB}^{A'B'})^T(s, t)~, \;\; 
\]
\[
(A_{AB}^{A'B'})^T(s, t) =\frac{s}{2\pi i} \int_{\delta -i\infty}^{\delta +i\infty}\frac{d\omega}{\sin(\pi\omega)}
\]
\be
\times \left((-1)^T\left(\frac{s}{s_0}\right)^\omega -\left(\frac{-s}{s_0}\right)^\omega\right)(A_{AB}^{A'B'})^T_\omega (t)~, 
\ee
through partial waves of amplitude $(F_{B'B})^{T}_\omega(k_\perp, q_\perp -k_\perp)$ of Reggeon scattering on the particle $B$
\be
(A_{AB}^{A'B'})^T_\omega (t) = \Phi^T_{A^{\prime}A}\int\frac{d^2 {k'}_\perp \; (F_{B'B})^{T}_\omega(k'_\perp, q_\perp -k'_\perp)}{({k'}_\perp^2-m^2)({(q-k')}_\perp^2-m^2)}
\ee
and  to write for them the equation   \cite{Fadin:1975cb}: 
\[
[2 + \omega - \alpha(k_\perp^2) - \alpha((q-k)_\perp^2)](F_{B'B})^{T}_\omega(k_\perp, q_\perp -k_\perp)=  \Phi^T_{B^{\prime}B}
+\frac{g^2}{(2\pi)^3}
\]
\[
\times\int \frac{d^2 {k'}_\perp \; (F_{B'B})^{T}_\omega(k'_\perp, q_\perp -k'_\perp)}{({k'}_\perp^2-m^2)({(q-k')}_\perp^2-m^2)}  \biggl[A_T(q_\perp^2) -\left(2-\frac12 T(T+1)\right)
\]
\be
\times \frac{(k_\perp^2-m^2)((q-k')_\perp^2-m^2) + ({k'}_\perp^2-m^2)((q-k)_\perp^2-m^2)}{{(k-k')}_\perp^2-m^2}   \biggr]~,  \label{preBFKL}
\ee
where $t=q^2 =q_\perp^2$,   $ \alpha(t)$  is the vector meson trajectory,
\be
\alpha(t) = 1+\frac{g^2}{(2\pi)^3}(t-m^2)\int \frac{d^2 k_\perp}{(k_\perp^2-m^2)((q-k)_\perp^2-m^2)}~, \;\; t=q^2 = q_\perp^2~, \label{trajectory-m}
\ee
$\Phi^T_{A^{\prime}A}$ and $\Phi^T_{B^{\prime}B}$ are the impact factor  for  the 
$A \rightarrow A^{\prime}$ and $B \rightarrow B^{\prime}$ transitions, 
\be
A_T(q^2) =\frac{m^2}{2}  + \left(2-\frac12 T(T+1)\right)(t -\frac12 m^2)~. \label{A_T}
\ee
It is not difficult to see that $\alpha(t)$ is equal to  $j_V(t)$  defined in (\ref{trajectory-m}).

It was shown in \cite{Fadin:1975cb} that there is bootstrap in the vector boson channel, corresponding to  $T=1$: the equation (\ref{preBFKL} ) gives in the $j$-plane only the same Regge pole which was assumed. Indeed,  the solution of Eq. (\ref{preBFKL}) for $T=1$ is 
\be
(F_{B'B})^{T}_\omega(k_\perp, q_\perp -k_\perp)\Big|_{T=1}=
\frac{\Phi^T_{B^{\prime}B}}{\omega +1-\alpha(q^2)}~,  \label{self1}
\ee
\be
(A_{AB}^{A'B'})^T_\omega (t)\Big|_{T=1}={\Phi^T_{A^{\prime}A}\Phi^T_{B^{\prime}B}}\Big|_{T=1}
\frac{(\alpha(q^2) -1)}{g^2(t-m^2)(\omega +1-\alpha(q^2))}~,  \label{self2}
\ee
and since $\Phi^T_{P^{\prime}P}\Big|_{T=1} = g\Gamma_{P^{\prime}P}$, where $\Gamma_{P^{\prime}P}$ is the vertex of reggeon-particle interaction, 
\be
(A_{AB}^{A'B'})(s, t)\Big|_{T=1}={\Gamma_{A^{\prime}A}\frac{\left((-u)^{\alpha(q^2)}-(-s)^{\alpha(q^2)}\right)}{(t-m^2)}{\Gamma_{B^{\prime}B}}~, \;\; u\approx -s
}~,  \label{self3}
\ee
in accordance with the Reggeization hypothesis.  Of course, this could  in no way be considered as a  proof of a hypothesis, because the hypothesis refers not only to elastic amplitudes. Nevertheless, it was the first step to the proof. Later the  bootstrap conditions for inelastic amplitudes  were formulated and the  proof of the hypothesis was  held on their basis \cite{Balitskii:1979}.  

So, a primary Reggeon in this approach (subsequently named BFKL) is the Reggeized gauge boson.   The Pomeron, which corresponds  to the rightmost $j$-plane singularity of the partial wave with $T=0$ in (\ref{preBFKL})  and determines the high energy behaviour of cross sections, appears  as  a compound state of two  Reggeized gauge bosons. 
It was shown \cite{Fadin:1975cb} that in the Pomeron  channel the leading $j$-plane singularity  turns out to be a square root branch point at 
\be
j=1 +N\frac{g^2}{\pi}\ln 2~, \label{jP}
\ee  
for the gauge group $SU(N)$.  Therefore,   the Froissart bound \cite{Froissart:1961ux} is violated as well as in QED \cite{Frolov:1970ij}, \cite{Cheng:1970xm}, although the mechanism of this violation  differs from QED:  in the non-Abelian theories cross sections for production of any fixed  number of particles decrease with energy due to the vector boson Reggeization, and the total cross section increases as power of $s$ only due to increasing number of opening channels.  The reason of the violation is that the $s$-channel unitarity 
is not fulfilled in the LLA. 

Another important observation made in \cite{Fadin:1975cb} was the growth of the typical  transverse momenta  with increasing $s$. It was conjectured that due to this growth and the asymptotic freedom only Regge poles may appear to the right from the point $j=1$.

Detailed consideration of the Pomeron channel was done in \cite{Kuraev:1977fs}. 

\subsection{Appearance of the BFKL and its development}
Appearance of the BFKL  should be attributed to  the end of the 1970s, 
when Lev turned to QCD and with his student Yan Balitsky  applied  the methods developed for non-Abelian theories with broken symmetry in \cite{Lipatov:1976zz}, \cite{Fadin:1975cb} -  \cite{Kuraev:1977fs}   to scattering of colourless particles in QCD. In 1978, a famous paper   \cite{Balitsky:1978ic} was published.    The results of \cite{Fadin:1975cb} (see (\ref{preBFKL}), (\ref{trajectory-m}))  can be applied in  the  massless limit (theory without spontaneous breaking of symmetry) only with some regularization (in the following dimensional regularization will be used). The reason is that they were obtained for scattering  of coloured particles (we will call them partons).  The key difference of colourless particles from partons is  that due to the gauge invariance   impact factors  of colourless particles, describing their interaction with Reggeized gluons, vanish at zero gluon momenta.  It was shown in \cite{Balitsky:1978ic} that in this case the transition to massless theory is not difficult: the equation   (\ref{preBFKL}) with  $\Phi^T_{B^{\prime}B}$  corresponding to the impact factor of a colourless particle is  free from the infrared singularities and therefore  can be used in QCD. 

From now on, we will talk about gauge theories with unbroken symmetry,  although, as it follows from the above, the  approach can be used  in theories with broken symmetry as well.  If not specified, QCD will be kept in mind, although for generality, the gauge group will be taken $SU(N_c)$, with the number of colours  $N_c$. 

An approximate theory can only be considered consistent when it is possible to calculate corrections to the approximation.  Development of the BFKL approach in the next-to-leading logarithmic approximation  (NLLA) was started in the late 1980s \cite{Fadin:1989kf}.  
In the NLLA, the same scheme of derivation of the BFKL equation as in the LLA is  applicable.  Again  only amplitudes with negative signature must be kept  on the right side of the unitarity relations (\ref{elastic discontinuity}), because leading terms in amplitudes with positive signature are imaginary and have one less degree of $\ln s$.  Moreover, only real parts of the amplitudes with negative signature are necessary because their imaginary parts are suppressed by one power of $\ln s$ compared to their real parts.   It was supposed that these amplitudes have the Regge pole form, i.e. are expressed in terms of the gluon trajectory and the Reggeon vertices.  Therefore it was necessary to calculate  one-loop corrections to the gluon trajectory and the vertices used in the LLA.  Also, instead of one gluon  two-gluon and quark-antiquark jets can be produced in the NLLA. Therefore  it was necessary to calculate new Reggeon vertices for the jet production. 

The calculations required a lot of effort and time. They started with the two-gluon jet production vertex \cite{Fadin:1989kf}. A lot of work was done and a lot of papers  were published (see \cite{Fadin:1989kf} - \cite{Fadin:1997hr}) before everything necessary for obtaining the  NLO BFKL kernel  in a colourless (Pomeron) channel at zero momentum transfer became available. Then this  kernel and his eigenvalues were  found \cite{Fadin:1998py} (see also \cite{Ciafaloni:1998gs}). It is necessary to notice here that although talking about  BFKL  kernel (or BFKL equation ) they usually mean exactly the kernel  for forward scattering, although the approach is also applicable to scattering with the transfer of both momentum  and colour. To date, a lot of work has been done and a number of important results on the development of the BFKL as applied to such processes have been obtained.

\section{\label{sec:basics}Basics of the BFKL approach}
\subsection{The gluon Reggeization}
The BFKL approach  is based on a remarkable property of QCD -- gluon Reggeization, which gives a very powerful tool for the description of high energy processes. The gluon Reggeization determines the form of QCD amplitudes at large energies and limited transverse momenta.  Due to the Reggeization,  dominant  amplitudes  have a simple factorized form in the  next-to-leading logarithmic approximation (NLLA)  as well as in the LLA.    The  Reggeization  allows  to express an infinite number of amplitudes   through several effective vertices and gluon trajectory. 

\begin{figure}[h]
\centerline{\includegraphics[width=5.5cm]{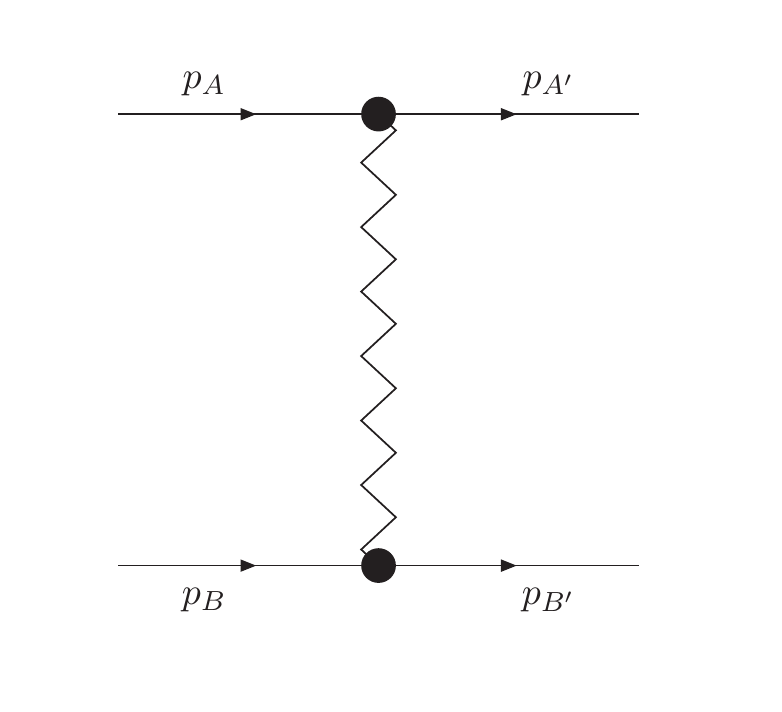}}
\caption{\label{2-2} Schematic representation of the process
 $A+B\rightarrow A'+B'$ with colour octet in the $t$ channel and negative signature. The zigzag line represents Reggeized gluon.}
\end{figure}

In the Regge kinematic
region  $ s \simeq - u \rightarrow \infty $,  $t$ fixed,   amplitudes of  the process $A+B \rightarrow A^{\prime }+B^{\prime }$ with a colour octet $t$-channel exchange and negative signature  can be depicted by the diagram of Fig.\ref{2-2} and have the Regge form  
 
\be
{{\cal A}}^{A' B'}_{AB} = {\Gamma}^c_{A^\prime
A} \left[\left({-s\over -t}\right)^{j(t)}-\left({s\over
-t}\right)^{j(t)}\right] \Gamma^c_{B^\prime B}~, \label{elastic Regge}
\ee
where  ${\Gamma}^c_{P^\prime P}$ are energy-independent particle-particle-Reggeon (PPR) vertices (or  scattering vertices), $c$ is a  colour index,   $j(t) =1 + \omega(t)$ is the  Reggeized gluon trajectory.

As it is seen from (\ref{elastic Regge}), in the leading order (LO) PPR vertices are determined by the  Born amplitudes, so that their calculation is trivial {\bf assuming the gluon Reggeization, i.e. the form (\ref{elastic Regge})}.  
Expressions for them are also quite  simple.  In the helicity basis
they have the same  form  for all partons (quarks and gluons in QCD):
\begin{equation}
 \Gamma _{P'P}^c\ =\ gT^c _{P'P}\delta_{\lambda_{P'}\lambda_P}\,,
\label{Gamma-c-helicity}
\end{equation}
where $T^c _{P^{\prime }P}$ are   matrix elements of the colour group
generators in the corresponding representations and $\lambda$ are parton helicities.
Except for a common coefficient the vertices (\ref{Gamma-c-helicity}) can be written down without calculation, because they are given by forward matrix
 elements of the conserved current. Note that Eq. (\ref{Gamma-c-helicity}) implies
 a definite choice  of the relative phase of spin wave functions
 of particles $P'$ and $P$. Evidently, the phase is zero at
 $t=0$. In (\ref{Gamma-c-helicity}) the $s$-channel helicity
 conservation is exhibited explicitly. Note that for
gluons and for massive quarks it is valid only in the LO. 
 
Again, as it is seen from (\ref{elastic Regge}),  the  LLA (one-loop)  trajectory  is determined by the $s$-channel
discontinuity of any   amplitude. It gives 
\[
\omega(t) =\frac{g^2N_c t}{2(2\pi)^{D-1}}\int\!
\frac{d^{D-2}q_1}{\q_1^{\:2}(\q-\q_1)^{2}}=-g^2 \frac{N_c \Gamma(1-\epsilon)}{(4
\pi)^{D/2}} \frac{\Gamma^2(\epsilon)}{\Gamma(2\epsilon)}(\q^{\:2})^\epsilon
\]
\be
\simeq-g^2
\frac{N_c \Gamma(1-\epsilon)}{(4 \pi)^{2+\epsilon}}
\frac{2}{\epsilon}(\q^{\:2})^\epsilon\,. \label{LO-omega}
\ee
In (\ref{LO-omega}) and below  the vector sign means transverse to the   $p_A, p_B$ plane  components,  $D=4-2\epsilon$ is the space-time dimension taken different from 4 to regularize infrared divergencies, which  are inevitable  in parton amplitudes. 
Of course, using the dimensional regularization in (\ref{trajectory-m}) at $m=0$
one obtains $\alpha(t) = 1+\omega(t)$, where $\omega(t)$ is given by (\ref{LO-omega}) 
at $N_c =2$.

\begin{figure}[h]
\centerline{\includegraphics[width=7.5cm]{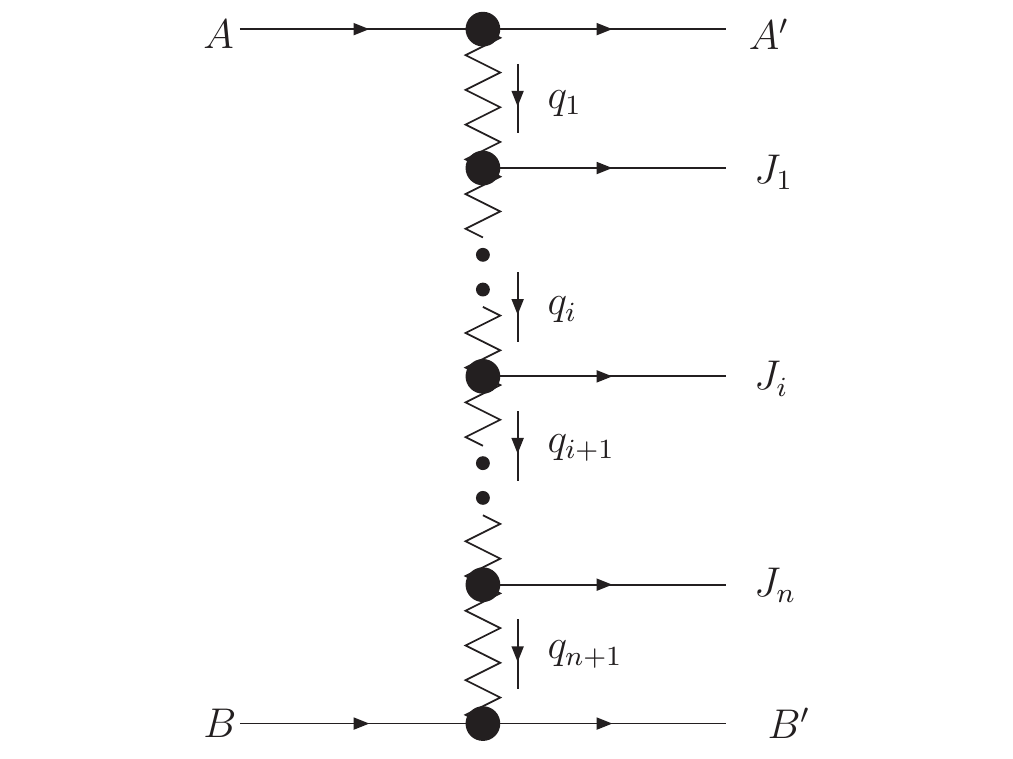}}
\caption{\label{2-2+n}  Schematic representation of the process $A+B\rightarrow A'+
J_1+\dots+J_{n}+ B'$ in MRK. The zig-zag lines represent Reggeized gluon exchange; the
black circles denote the Reggeon vertices; $q_i$ are Reggeon momenta.}
\end{figure}

The Reggeization means also definite (multi-Regge) form   of  production  amplitudes in the multi-Regge kinematics  (MRK). MRK is the kinematics  where all particles have
limited (not growing with $s$) transverse momenta and are combined into jets with limited invariant mass of each jet and large (growing with $s$) invariant masses of any pair of the jets.  This kinematics gives dominant contributions to cross sections of QCD processes at high energy $\sqrt s$. In  each order of perturbation theory dominant (having the largest $\ln s$ degrees) are amplitudes with gluon quantum numbers and a negative signature. For the amplitude  ${\cal A}_{2\rightarrow n+2}$ of  the  process $A+B\rightarrow A'+J_1+\ldots+J_n+B'$  of production of $n$ jets with momenta $k_1, k_2, \ldots k_n $
 the  MRK means
 \be
 s \gg s_i \gg |t_i|\simeq \qs_i,~~~~ \;\; s\simeq \frac{ \prod_{i=1}^{n+1}s_i} {\prod_{i=1}^{n}(\vks_i+k_i^2)}~, \label{MRK}
 \ee
 where
 \be
 s=(p_A+p_B)^2, \; s_i=(k_{i-1}+k_i)^2, \;\;i=1, \cdot\cdot\cdot n+1, \;   \; k_{0}\equiv P_{A'}, \;\;k_{n+1}\equiv P_{B'},\label{notation1}
 \ee
 \be
 q_1=p_A-p_A',\;   q_{j+1} = q_{j}-k_j, \; j=1, \cdot\cdot\cdot n, \; \; q_{n+1} =p_{B'} -p_B~.   \label{notation2}
 \ee
In this region  the amplitudes  ${\cal A}_{2\rightarrow n+2}$ can be represented by
Fig. \ref{2-2+n}. 

\noindent 
Multi-particle amplitudes  have a complicated
analytical structure. They are not simple even  in MRK (see, for instance,
 \cite{Bartels:1974tj} - \cite{Fadin:1993wh}).
Fortunately, only real parts  of these amplitudes are used in the 
BFKL approach in the NLLA as well as in the LLA.  The reason is that imaginary parts of amplitudes with negative signature are suppressed by one power of $\ln s$ compared to their real parts, so that account of them in the   right part of the unitarity relations (\ref{elastic discontinuity})  means  loss of two powers  of $\ln s$  (as well as account of amplitudes with positive signature).  Restricting
 ourselves to the real parts  we can  write  
\[ 
 \Re {\cal A}_{2\rightarrow n+2}=2s\Gamma^{{R}_1}_{
 A' A} \left( \prod_{i=1}^{n}
  \frac{1}{t_{i}}\Big(\frac{s_i}{|\vk_{i-1}||\vk_{i}|}\Big)^{\omega(t_i)}
 \gamma^{G_i}_{R_i{R}_{i+1}}\right)
\] 
\be 
 \frac{1}{t_{n+1}}\Big(\frac{s_{n+1}}{|k_{n}||\q_{n+1}|}\Big)^{\omega(t_{n+1})}
 \Gamma^{{R}_{n+1}}
 _{B' B}~.
 \label{A 2-2+n}
 \ee
 Here ${\Gamma}^{{R}}_{ A'A}$ and $\Gamma^{{R}} _{B' B}$ are the same scattering vertices  as in (\ref{elastic Regge})    and  $\gamma ^{J_i}_{{R}_i {R}_{i+1}}$ are the  Reggeon-Reggeon-Jet (RRJ) vertices (or production vertices), i.e. the effective vertices for production of jets $ J_i$ with momenta
 $k_{i}$=$q_{i}-q_{i+1}$ in collisions of Reggeons with momenta $q_{i}$ and $-q_{i+1}$.  
 
In the LLA only gluons  can be produced, $J=G$,   
\be
\gamma ^{G}_{{R}_1 {R}_{2}} =g T^c_{c_1 c_2} e^*_\mu(k) C^\mu(q_2, q_1) \,,
\label{gamma-RRG}
\ee  
where $k=q_1-q_2$,  $e_\mu(k)$  and $c$  are the gluon momentum, polarization vector and colour index, $c_1$ and $c_2$ are the colour indices of the Reggeons ${R}_1$ and ${R}_1$ respectively,  and  $C^\mu(q_1, q_2)$ is the famous Lipatov's vertex \cite{Lipatov:1976zz}:
\[
 C_{\mu}(q_2,q_1)\ =\ -q_{1\mu}-q_{2\mu}+p_{1\mu}
\left(\frac{q_1^2}{kp_1}+2\frac{kp_2}{p_1p_2}\right) -
p_{2\mu}\left(\frac{q_2^2}{kp_2}+2\frac{kp_1}{p_1p_2}\right)
\]
\be
 =\ -q_{1\perp\mu}-q_{2\perp\mu}-\frac{p_{1\mu}}{2(kp_1)}
\left(k_{\perp}^2-2q_{1\perp}^2\right)+\frac{p_{2\mu}}{2(kp_2)}
\left(k_{\perp}^2-2q_{2\perp}^2\right). 
\label{vector-C}
\ee
The Reggeon vertices and the gluon trajectory are known now  in the next-to-leading order (NLO), that means the one-loop approximation for the vertices and the two-loop approximation for the trajectory. It is just the accuracy which is required for the derivation of the BFKL equation in the  NLLA.
Validity of the forms  (\ref{elastic Regge}) and (\ref{A 2-2+n}) is proved now   in all orders of perturbation theory  in the coupling constant $g$ both  in the LLA~\cite{Balitskii:1979} and in the  NLLA~\cite{Fadin:2006bj,Kozlov:2011zza,Kozlov:2012zz,Kozlov:2012zza,Fadin:2015zea}. 

In the LLA only gluons  can be produced. In the NLA  one has to consider not only the  amplitudes~(\ref{elastic Regge}),  (\ref{A 2-2+n}),   but also amplitudes obtained from them by  replacement of one of final particles by  a  jet containing a couple of particles with fixed (of order of transverse momenta) invariant  mass.

\subsection{The scheme of derivation of the BFKL equation}
\begin{figure}[h]
 \centerline{\includegraphics[width=7.5cm]{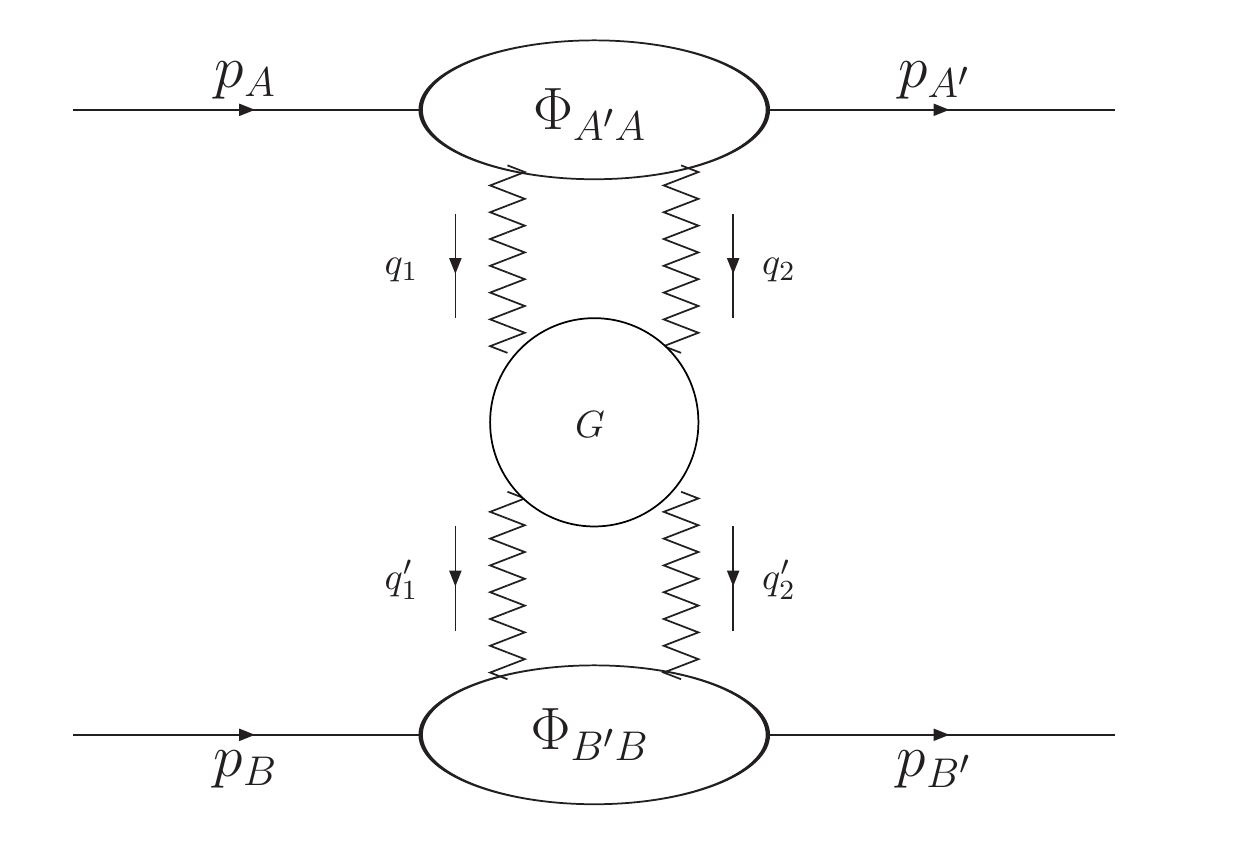}}
 \caption{\label{discontinuity} Schematic representation of the $s$-channel discontinuities of amplitudes  $A+B\rightarrow A'+B'$.}
 \end{figure}

The Reggeization provides a simple derivation of the BFKL equation both in the LLA and NLLA. Two-to-two scattering amplitudes  with all possible quantum
numbers in the $t$--channel are calculated using the  amplitudes ~ (\ref{A 2-2+n}) in the  $s$-channel unitarity  relations (\ref{elastic discontinuity}) and analyticity.
The $s$-channel discontinuities  of the  amplitude  ${\cal A}_{AB}^{A'B'}$  for the high energy process $A+B \rightarrow A'+B'$ at fixed momentum transfer $\sqrt{-t}$  may be presented by Fig.\ref{discontinuity},
where $q_1+q_2=q'_1+q'_2=q=p_A-p_{A'}=p_{B'}-p_B$ and 
symbolically written as the convolution
\[\;\;\;\;\;\;\;\; \Phi _{A^{\prime }A}\;\otimes \;G\;\otimes \;
\Phi_{B^{\prime }B}
\]
where the impact factors $\Phi _{A^{\prime}A}$ and  $\Phi _{B^{\prime }B}$ describe  transitions $A\rightarrow A^{\prime}$ and  $B\rightarrow B^{\prime }$ due to interactions  with Reggeized gluons, $G$ is the  Green's function for two
interacting Reggeized gluons with  an operator form 
\be
{\hat{\cal
G}}=e^{Y\hat{\cal{K}}}~, \label{G}
\ee
where $\;Y=\ln(s/s_0)$, $s_0$  is an energy scale, ${\hat{\cal{K}}}$ is the  BFKL kernel. The impact factors and the BFKL kernel   are expressed in terms of the Reggeon vertices and trajectory. Energy dependence of scattering amplitudes is determined by the  BFKL kernel, which is universal (process independent). The kernel \be 
{\hat{\cal {K}}}={\hat{{\omega}_1}}+{\hat{{\omega}}_2}+ {\hat{\cal{K}}_r} \label{K}
\ee
is expressed through the Regge trajectories ${\hat{{\omega}}_1}$ and ${\hat{{\omega}}_2}$ of two gluons  and the ``real part" ${\hat{\cal {K}}_r}$ describing production of particles in their interaction: 
\be 
{\hat{\cal {K}}_r} ={\hat{\cal {K}}_G}+{\hat{\cal {K}}_{Q\bar Q}}+{\hat{\cal {K}}_{GG}} ~.  \label{Kr}
\ee
In the LLA only  ${\hat{\cal {K}}_G}$ must be kept, because only  gluons can be produced; in the NLLA production of quark-antiquark (${Q\bar Q}$)  and gluon ($GG$) pairs  is  also possible.
The impact factors $\Phi _{A^{\prime}A}$ and $\Phi _{B^{\prime }B}$ describing  transitions $A\rightarrow A^{\prime }\;\;\;$ and  $\;\;\;B\rightarrow
B^{\prime }$ depend on  properties of scattering particles.
All energy dependence is contained in the Green's function $G$ for two interacting Reggeized gluons.

While the contribution of the trajectories to the kernel  is diagonal in colour, the contribution associated with the production of real particles  can change colour states. Using the projection operators $\hat{\cal P}^{R}$ on the representation $R$ of the colour group, one  can perform the decomposition 
\be
{\hat{\cal {K}}} = \sum_{R}\hat{\cal P}^{R} \hat{\cal {K}}^R ~. 
\ee
In QCD, possible representations are  $\bf{1}$, $\bf{8}_a$, $\bf{8}_s$, $\bf{10}$,  $\bf{10^*}$ and  $\bf{27}$. 
In the following, we will consider only the singlet representation, corresponding to 
scattering of colourless particles, and omit the index $R$. 

\subsection{Forward BFKL kernel}
Separate contributions to the kernel  (\ref{K}) are singular. They were calculated in the dimensional regularization, so that in \cite{Fadin:1998py}  the kernel also was written at  the space-time dimension $D=4-2\epsilon$, different from the physical one. To work with this kernel is rather  difficult. But the kernel is infrared safe, i.e. singularities in various parts cancel. It makes possible to perform the cancellation of the most singular terms and to write the kernel in the form \cite{Fadin:2009za} (here the Reggeon momenta are q and l )
\[
{\cal K}(\boq, \bol) =\frac{\alpha_{s}(\mu^2)N_{c}}{2\pi^{2}}\left[\frac{2}
{(\boq-\bol)^{2}}-\delta(\boq -\bol)\int \frac{d\bok\;\boq^{\;2}}{(\boq-\bok)^{2}
\bok^{\:2}}\right]
\]
\[
\times \left[1+\frac{\alpha_s(\mu^2)N_{c}}{4\pi} \left(\frac{67}{9}-2\zeta(2)
-\frac{10}{9}\frac{n_f}{N_c}\right)\right]
\]
\[
+\frac{\alpha^2_{s}(\mu^2)N^2_{c}}{4\pi^{3}}\left[\frac{1}{(\boq
-\bol)^2}\left(\frac{\beta_0}{N_c} \ln\left( \frac{{\mu}^2}{(\boq-\bol)^{\;2}}
\right)-\ln^2\left(\frac{\boq^{\;{2}}}{\bol^{\; 2}}\right)\right) +f_1(\boq, \bol) \right.
\]
\begin{equation}
\left.
+
f_2(\boq, \bol)+\delta(\boq-\bol)\left( \frac{\beta_0}{2N_c}\int
\frac{d\bok\;\boq^{\;2}}{(\boq-\bok)^{2}\bok^{\:2}}
\ln\left(\frac{(\boq-\bok)^{2}\bok^{\:2}}{{\mu}^2 \boq^{\;2}} \right)+6\pi\zeta(3)\right)
\right]~, \label{forward K at D=4}
\end{equation}
where 
where  $\beta_0=11 N_c/3-2n_f/3$ is the first coefficient of the $\beta$-function,
\[
f_1(\boq_1, \boq_2) \ =  -\ \frac{2(\boq_1^{~2}-\boq_2^{~2})}
{\vks(\boq_1+\boq_2)^2} \Bigg(\frac{1}{2}
\ln \left(
\frac{{\boq_1}^{~2}}{{\boq_2}^{~2}}\right) 
\ln\left(\frac{{\boq_1}^{~2}{\boq_2}^{~2}\bok^{\:4}
}{\left({\qs_1}+{\qs_2}\right)^4}\right) +Li_2\left( -\frac{{\boq_1}^{~2}
}{{\boq_2}^{~2}}\right)
\]
\begin{equation}
-Li_2\left( -\frac{{\boq_2}^{~2}}{{\boq_1}^{~2}} \right) \Bigg)-\left(
1-{(\boq_1^{~2}-\boq_2^{~2})^2}/{\bok^{\:2}(\boq_1+\boq_2)^2}\right) \left(
\int\limits_0^1-\int\limits_1^\infty \right) \frac{dz\,\ln
\frac{(z{\boq_1})^2}{({\boq_2})^2}}{(\boq_2-z\boq_1)^2}, \label{f 1 q 1 q 2}
\end{equation}
\[
f_2(\boq_1, \boq_2) \ =  -\  \left(1+\frac{n_f}{N_c^3}\right)
\frac{2{\boq_1}^{~2}{\boq_2}
^{~2}-3({\boq_1}{\boq_2})^2}{16{\boq_1}^{~2}{\boq_2}^{~2}}\left( \frac
2{{\boq_2}^{~2}}+\frac 2{{\boq_1}^{~2}}+\Big(\frac 1{{\boq_2}^{~2}} -\frac
1{{\boq_1}^{~2}}\Big)\ln \frac{{\boq_1}^{~2}}{{\boq_2}^{~2}}\right)
\]
\[
-\ \left(\! 3+\Big(1+\frac{n_f}{N_c^3}\Big)\!\left(\!1
-\frac{({\boq_1}^{~2}+{\boq_2}^{~2})^2}{8{\boq_1}^{~2}{\boq_2}^{~2}}
-\frac{2{\boq_1}^{~2}{\boq_2}^{~2}-3{\boq_1}^{~4}-3{\boq_2}^{~4}}
{16{\boq_1}^{~4}{\boq_2}^{~4}}({\boq_1}{\boq_2})^2 \!\right)\! \right)
\]
\begin{equation}
\times\ \int\limits_0^\infty \frac{dx\,\ln \left| \frac{1+x}{1-x}\right|
}{{\boq_1}^{~2}+x^2{\boq_2}^{~2}}~,  \label{f 2 q 1 q 2}
\end{equation}
where $\bok =\boq_1-\boq_2~.$

This representation greatly simplifies the calculation of the eigenvalues of the kernel.  Strictly
speaking, because of the charge renormalization the eigenfunctions of the LO kernel  
$\left({\bol^{\:2}}\right)^{\gamma-1}$, which at $\gamma = 1/2 + i\nu, \;\; -\infty <\nu < \infty $ form a complete set  for the kernel averaged over angles,  
are  not any more eigenfunctions of the NLO kernel. But with the NLO accuracy one  can write
\begin{equation}
\alpha_s(\mu^2)=\alpha_s(\boq^{\:2})\left(1+\beta_0\frac{\alpha_s(\boq^{\:2})}{4\pi}
\ln\left(\frac{\boq^{\:2}}{\mu^2}\right)\right)~,
\end{equation}
and obtain  
\[
\int d^2l\;{\cal K}(\boq,\bol)\left(\frac{\bol^{\:2}}{\boq^{\:2}}\right)^{\gamma-1} =
\omega(\boq^{\,2}, \gamma)= \frac{\alpha _s(\boq^{~2})\,N_c\,}\pi \chi(\gamma )~,\;\;\
\]
\begin{equation}
\chi(\gamma)= \chi_B (\gamma )+ \frac{\alpha _sN_c}{\pi } \chi^{(1)} (\gamma )~.
\label{chi-function}
\end{equation}
Here $\chi_B(\gamma ) $ gives the LO eigenvalues, 
\be
\chi_B(\gamma) = 2\psi(1) - \psi(\gamma)- \psi(1-\gamma)
\ee
and  the correction
$\chi^{(1)}(\gamma)$ is:
\[
\chi^{(1)}(\gamma)\ =\ -\frac14 \Bigg[\frac{\beta_0}{2N_c}\left(\chi_B^2(\gamma)(\gamma)
+\chi^{\prime}_B(\gamma)-\ 6\zeta (3) \right)
\]\
\[
+\frac{\pi ^2\cos(\pi \gamma )}{\sin^2(\pi \gamma )(1-2\gamma )}\left( 3+\left(
1+\frac{n_f}{N_c^3}\right) \frac{2+3\gamma (1-\gamma )}{(3-2\gamma )(1+2\gamma )}\right)
\]
\begin{equation}
-\ \left( \frac{67}9-\frac{\pi ^2}3-\frac{10}9\frac{n_f}{N_c}\right) \chi_B(\gamma)+\
\chi_B^{\prime\prime}(\gamma))- \frac{\pi ^3}{\sin (\pi \gamma )}+4\phi (\gamma )\Bigg]~,
\label{chi-correction}
\end{equation}
where
\[
\phi (\gamma) = -\int\limits_0^1\frac{dx}{1+x}\left( x^{\gamma-1} +x^{-\gamma}\right)
\int_x^1\frac{dt}t\ln (1-t)
\]
\be
= \sum_{n=0}^\infty (-1)^n\left[ \frac{\psi (n+1+\gamma )-\psi (1)}{(n+\gamma
)^2}+\frac{\psi (n+2-\gamma )-\psi (1)}{(n+1-\gamma )^2}\right]. \label{phi-function}
\ee
This result was firstly obtained in \cite{Fadin:1998py}. 

The representation (\ref{forward K at D=4}) is useful also for finding all eigenvalues of the kernel.  Defining 
them as 
\[
\int d^2l\;{ K}(\boq,\bol)\left(\frac{\bol^{\:2}}{\boq^{\:2}}\right)^{\gamma-1}
e^{in(\phi_l-\phi_q)} 
\]
\be
= \frac{\alpha _s(\boq^{~2})\,N_c\,}{\pi} \left(\chi_B(\gamma,
n)+\frac{\alpha_sN_c}{\pi}\chi^{(1)}( \gamma, n)\right)~, \label{omega(n, gamma) NLO}
\ee
where 
\be
\chi_B(\gamma,n) = 2\psi(1) - \psi(\gamma+\frac{|n|}{2})-\psi(1-\gamma+\frac{|n|}{2})~,  \label{chiBgamman}
\ee 
we come to the correction obtained in \cite{Kotikov:2000pm}:
\[
4\chi^{(1)}\left(\gamma, n\right)  \ = \   -\frac{\beta_0}{2N_c}\left(\chi_B^{2}(\gamma,
n)+\chi_B^{\prime}(\gamma,  n) \right)+6\zeta\left(  3\right) -
\chi_B^{\prime\prime}(\gamma, n)
\]
\begin{equation}
+\left( \frac{67}9-\frac{\pi ^2}3-\frac{10}9\frac{n_f}{N_c}\right) \chi_B(\gamma,
n)-2\Phi\left(  n,\gamma\right) -2\Phi\left(  n,1-\gamma\right)  +F\left(
n,\gamma\right)~,  \label{chi1gamman}
\end{equation}
where
\[
F(n,\gamma)=\frac{\pi^{2}\cos\left(  \pi\gamma\right)  }{\sin^{2}\left( \pi \gamma\right)
\left( 1-2\gamma\right)  }\Biggl[  \frac{\gamma\left( 1-\gamma\right)  \left(
\delta_{n,2}+\delta_{n,-2}\right)  }{2\left( 3-2\gamma\right)  \left(  1+2\gamma\right)
}
\]
\begin{equation}
\times\left(1+\frac{n_f}{N_c^3}\right)-\left(  \frac{3\gamma\left( 1-\gamma\right)
+2}{\left(  3-2\gamma\right)  \left(  1+2\gamma\right)
}\left(1+\frac{n_f}{N_c^3}\right)+3\right)  \delta_{n,0}\Biggr]~,\label{F(n,gamma)}
\end{equation}
and
\[
\Phi\left(  n,\gamma\right)  =\int_{0}^{1}\frac{dt}{1+t}t^{\gamma -1+n/2}\left\{
\frac{\pi^{2}}{12}-\frac{1}{2}\psi^{\prime}\left(  \frac
{n+1}{2}\right)  -{Li}_{2}\left(  t\right)  \right.
\]%
\[
   -{Li}_{2}\left(  -t\right) -\left(  \psi\left(  n+1\right)  -\psi\left(  1\right)  +\ln\left( 1+t\right)
+\sum_{k=1}^{\infty}\frac{\left(  -t\right)  ^{k}}{k+n}\right) \ln t
\]
\be
\left.-\sum_{k=1}^{\infty}\frac{t^{k}}{\left(  k+n\right)  ^{2}}\left[ 1-\left(  -1\right)
^{k}\right]  \right\}~. \label{Phi(  n,gamma)}
\ee

\section{Investigation of the BFKL properties}
In the wide range of problems that Lev dealt with, the BFKL approach was one of the first. Lev many times turned to it, and his analysis was always distinguished by its depth and originality.
\subsection{Conformal invariance}
 In 1985 he  discovered  that for scattering of colourless particles the LO BFKL equation can be solved in  a general form not only for the forward case,  but for general momentum transfer \cite{Lipatov:1985uk}. This striking fact is related to the remarkable property of the BFKL equation discovered in \cite{Lipatov:1985uk}: for  scattering of colourless particles it can be written in the special representation (which was called later M\"{o}bius representation \cite{Bartels:2004ef}),  which is invariant with respect to  conformal (M\"{o}bius)  transformations   in the impact parameter space  $\vec r_{1}, \vec
r_{2}$ of the Reggeized gluons. In this representation the BFKL kernel  $\hat{\cal{K}}_M$ operates in the space of functions having the  colour transparency property (turning into  zero at $\vec r_{1}=\vec r_{2}$).  Note that the colourlessness means the Pomeron exchange (while  the reverse statement is  not correct)  and that  it  is necessary for writing  the BFKL equation in the M\"{o}bius representation.  The colourlessness of colliding particles provides the gauge invariance  of  impact factors, i.e. their vanishing  at zero  Reggeized gluon momenta. Together with an analogous property of the BFKL kernel, it gives a possibility to restrict  the space of functions where the kernel acts by the functions having the  colour transparency.

The conformal invariance of the BFKL equation in the M\"{o}bius representation is extremely important because it permits to classify  all solutions of the homogeneous equation and to find their complete set.  Another significant property  is the holomorphic separability of  BFKL kernel in this representation. Actually,  this property was used already in \cite{Lipatov:1985uk}; it was explicitly exhibited  and  used later \cite{Lipatov:1990zb, Lipatov:1993qn} for solving the Bartels-Kwiecinski-Praszalowichz (BKP) equation  -- the  generalization  of  the BFKL equation  to the case of many Reggeized gluons in the $t$--channel \cite{Bartels:1980pe, Kwiecinski:1980wb}. 
The conformal invariance permits to obtain  eigenfunctions of the kernel $\hat{\cal{K}}_M$ from eigenfunctions for forward scattering by   transformations of   the M\"{o}bius group  (inversion and shift).  In   \cite{Lipatov:1985uk} they were
chosen in the form
\begin{equation}
E_{n,\nu}(\bor_{10}, \bor_{20})= (-1)^n\left(\frac{z_{12} }{z_{10}
z_{20}}\right)^{\frac12+i\nu-\frac{n}{2}} \left(\frac{z^*_{12} }{z^*_{10}
z^*_{20}}\right)^{\frac12+i\nu+\frac{n}{2}}~, \label{E n nu}
\end{equation}
where $z= x+iy$ for $\bor = (x, y)$, the vector $\bor_0$ is introduced for indexing of the wave functions,
$\bor_{ij}=\bor_i - \bor_j$, $n$ is integer and $\nu$ is real. The eigenvalues coincide
with the eigenvalues of the forward kernel 
\be 
\omega_B(\gamma, n) = \frac{\alpha_s N_c}{\pi}\chi_B(\gamma,n) ~, \gamma =1/2+i\nu~, 
\ee
$\chi_B(\gamma,n)$ is given in (\ref{chiBgamman}).   One can see it noticing that the eigenvalues do not depend on
$\bor_0$ and that integration over $\bor_0$ gives
\begin{equation}
\int \,d\bor_0\,E_{n,\nu}(\bor_{10}, \bor_{20})= C_{n,
\nu}\,(z_{12})^{\frac12-i\nu+\frac{n}{2}}(z^*_{12})^{\frac12-i\nu- \frac{n}{2}} =C_{n,
\nu}\,e^{i\,n\phi_{12}}\bor_{12}^{2(1-\gamma)} ~, \label{intE n nu}
\end{equation}
where $\phi_{12}$ is the azimuthal angle of the vector $\bor_{12}$. Then the problem is
reduced to the calculations of the integrals for  the
forward case. The form of the integral (\ref{intE n nu}) can be easily obtained using
the change of the integration variables $z_{10}=z_{12}z$; the coefficient $C_{n, \nu}$ is
irrelevant for the calculation of the eigenvalues.  In fact, it was found in
\cite{Lipatov:1985uk}:
\begin{equation}
C_{n, \nu}=\frac{\pi 2^{4\,i\,\nu}}{|n|-2\,i\,\nu}\frac{\Gamma(\kappa)}
{\Gamma(\kappa^*)}\frac{\Gamma(\kappa^*-1/2)} {\Gamma(\kappa-1/2)}~, \;\;
\kappa=\frac{1}{2}-i\nu +\frac{|n|}{2}~.
\end{equation}

One can easily show that
\[
z^2_{12}\frac{\partial}{\partial z_1}\frac{\partial}{\partial z_2}\,E_{n,\nu}=
\lambda_{n,\nu}E_{n,\nu}~, \;\;z^{*\,2}_{12}\frac{\partial}{\partial
z^*_1}\frac{\partial}{\partial z^*_2}\,E_{n,\nu}= \lambda_{n,-\nu}E_{n,\nu}~,
\]
\begin{equation}
\lambda_{n,\nu}=\frac{1}{4}-(\frac{n}{2}-i\nu)^2~. \label{derivatives E}
\end{equation}
The completeness condition derived in \cite{Lipatov:1985uk} has the form
\[
4\sum_{n=-\infty}^{+\infty}\int_{-\infty}^{+\infty} \ d\nu \int \ d^2r_0
\frac{n^2+4\nu^2}{\bor^2_{12}\bor^2_{1'2'}}E_{n,\nu}(\bor_{10},
\bor_{20})E^*_{n,\nu}(\bor_{1'0}, \bor_{2'0})
\]
\begin{equation}
=(2\pi)^4\delta(\bor_{11'})\delta(\bor_{22'})~. \label{E n nucompleteness}
\end{equation}
Using (\ref{derivatives E}) and (\ref{E n nucompleteness}) one obtains  for the Green's function in the impact parameter space
\[
\langle\bor_{1},\bor_{2}|\frac{1}{\omega-\hat{\cal{K}}_M}\frac{1}
{\hat{\boq}_{\,1}^{\,2}\hat{\boq}_{\,2}^{\,2}}|\bor_{\,1}^{\;\prime} ,\bor_{\,2}^{\;\prime}\rangle
\]
\begin{equation}=4\sum_{n=-\infty}^{+\infty}\int_{-\infty}^{+\infty} \ d\nu \int \ d^2r_0
\frac{({n^2+4\nu^2})E_{n,\nu}(\bor_{10},
\bor_{20})E^*_{n,\nu}(\bor_{1'0},\bor_{2'0})}{[(n+1)^2+4\nu^2][(n-1)^2+4\nu^2]
(\omega-\omega(\gamma, n))}~.
\end{equation}
For $n=\pm 1$ the integral over $\nu$ here must be taken in the sense of its principal
value.

Later  an explicit expression for the kernel ${\cal{K}_M}$  in the impact parameter space was obtained \cite{Fadin:2006ha}: 
\[
\langle\bor_{1},\bor_{2}|\hat{\cal{K}}_M|\bor_{1}^{\;\prime}
,\bor_{2}^{\;\prime}\rangle
\]
\begin{equation}
=\frac{\alpha_s N_c}{2\pi^2}\int\,{d\bor_0 }\
\frac{\bor_{12}^{\,2}}{\bor_{10}^{\,2}\bor_{20}^{\;2}} \left(\delta(\bor_{11'})
\delta(\bor_{02'})+\delta(\bor_{22'})\delta(\bor_{01'})-\delta(\bor_{11'})
\delta(\bor_{22'}) \right)~, \label{Mobius LO}
\end{equation}
where the subscript $M$ denotes the M\"{o}bius  form.  It is worth to note that the
ultraviolet singularities of separate terms in  (\ref{Mobius LO})  cancel in their sum
with account of the dipole property of the target impact factors.

The transformations of the  M\"{o}bius group in the two-dimensional space $\bor=(x,y)$
can be written as
\begin{equation}
z\rightarrow \frac{az+b}{cz+d}\;, \;\;\;
\end{equation}
where $z=x+iy, \;\; a,b,c,d$ are complex numbers, with $ad-bc\neq 0$. Under these
transformations, one has
\[
z_1-z_2\rightarrow \frac{z_1-z_2}{(cz_1+d)(cz_2+d)}(ad-bc)\;,
\]
\begin{equation}
dzdz^*\rightarrow dzdz^*\frac{|ad-bc|^2}{\left|(cz+d)^2\right|^2}\;,
\end{equation}
so that the conformal invariance of (\ref{Mobius LO}) is evident.

It turns out that the M\"{o}bius form (\ref{Mobius LO})  coincides  with the kernel of the colour dipole model~\cite{Nikolaev:1993th}-\cite{Mueller:1994jq} formulated in the coordinate space. It means that the colour dipole  kernel  could have been  written by Lev  as early as 1985.  It was not done only by chance, because he used the operator form of the kernel instead of the explicit expression for it in the impact parameter space.

Investigation of the conformal properties of the BFKL kernel in the NLO \cite{Fadin:2007ee} - \cite{Fadin:2009gh}   showed later that  the ambiguity of the NLO kernel  allows to present the kernel in the form where the conformal invariance is violated only by renormalization.

In the same paper \cite{Lipatov:1985uk}  Lev performed another great study: account of running of the coupling constant. Strictly speaking, this account  oversteps the limits of the LLA. Nevertheless, it is not unreasonable to take it into account just in the LO  BFKL kernel in order to understand qualitative effects of the running. 
One important  effect  discovered by Lev is the conversion of the cut
 $[0, \omega_P^B]$ into an infinite series of moving poles with limiting
 point at $\omega=0$.  The Pomeron  trajectories found  in \cite{Lipatov:1985uk} are represented in the form
\begin{equation}
 \omega_k(t)=\frac{c}{k+\eta(q^2)+1/4}~, \;\; k=0,\, \pm 1,\, \pm 2\, ...,
\end{equation}
where  $c$ is a calculable constant. The function $\eta(q^2)$ is determined by large
distances, but  is limited by the inequalities $-1\leq 4\eta \leq 1$ at small $|t|$.  In
the region of large $|t|$ ($\alpha_s(|t|)\ll 1$) the family of poles is  approximated by
the moving cut with the branch point
\begin{equation}
 \omega(t)=\frac{4N_c}{\pi}\ln 2 \,\,\alpha_s(|t|)~.
\end{equation}

\subsection{Relation with DGLAP}
The relation between the eigenvalues of BFKL kernel $\frac{\alpha_s N_c}{\pi}\chi(\gamma)$ and the anomalous dimension $\gamma_\omega$ of the twist-2 operators  with $j=1+\omega$ was first discovered in  the LLA in \cite{Jaroszewicz:1982gr}, where it  has the form: 
\be
\omega = \frac{\alpha_s N_c}{\pi}\chi(\gamma_\omega)~. 
\ee
It is a remarkable result because  it means that the BFKL approach permits to obtain  the resummation of the most singular at $j=1$ terms  in the anomalous dimension of the twist-two operators.

It was noticed in \cite{Fadin:1998py} that in the NLLA this relation must be changed. 
The reason is that  energy scale $s_0$ becomes significant in the NNLLA, and this scale is different from DGLAP.  In the NLA   $s_0=Q_AQ_B$, where $Q_A$ and
$Q_B$ are typical momenta for the impact factors $\Phi_A$ and $\Phi_B$.  Therefore 
the deep-inelastic moments  are defined as
\begin{equation}
M_\omega(Q^2)\ =\ \frac1{4(2\pi)^2}\int\frac{d\boq}{\boq^2}
\int\frac{d\bol}{\bol^2}\Phi_A(\boq)e^{\omega\ln(Q/Q_B)}G_\omega(\boq, \bol)
\Phi_B(\bol)\,.  \label{M omega NLO}
\end{equation}
As the result, for the anomalous dimension
$\gamma_\omega(\alpha_s)$ one has 
\begin{equation}
\gamma_\omega =\frac{\omega}{2} + \chi^{-1}\left(\frac{\pi\omega}{N_c\alpha_s}\right)~,
\label{gamma omega from omega}
\end{equation}
where $\chi^{-1}$ is the inverse function. In other words, the anomalous dimensions
$\gamma_{\omega}(\alpha_s)$ of the twist-2 operators near the point $\omega =0$ are
determined from the solution of the equation
\[
\omega\ =\ \frac{\alpha _sN_c}{\pi}\chi (\gamma -\frac{\omega}{2} ) \simeq \frac{\alpha
_sN_c}{\pi} \chi(\gamma)-\frac{\alpha^2_sN^2_c} {2\pi^2}\chi(\gamma)\chi'(\gamma)
\]
\[
\simeq \frac{\alpha _sN_c}\pi
 \frac 1\gamma  -\frac{\alpha _s^2 N_c^2}{
4\,\pi ^2}\left( \frac{11+2n_f/N_c^3}{3\,\gamma ^2}+\ \frac{n_f(10+13/N_c^2)}{9\gamma
\,N_c}\right.
\]
\begin{equation}
\left. +\frac{395}{27}-2\zeta (3)- \frac{11}3\frac{\pi
^2}6+\frac{n_f}{N_c^3}\left(\frac{71}{27}-\frac{\pi ^2} 9\right)+{\cal O}(\gamma)\right)
\label{omega NLO from gamma}
\end{equation}
for $\gamma \rightarrow 0$.   This equation was derived in \cite{Fadin:1998py} and used
to  reproduce the known results and predict the higher loop correction for $\omega
\rightarrow 0$:
\begin{eqnarray}
 \gamma &\simeq& \frac{\alpha _s\,N_c}\pi \left(\frac 1\omega
 -\frac{11}{12}-\frac{n_f}{6N_c^3}\right)-\left( \frac{\alpha _s\,}\pi \right)
 ^2\frac{n_f\,N_c}{6\,\omega} \left(\frac 53+\frac{13}{6N_c^2}\right)
 \nonumber\\
&&-\ \frac1{4\omega ^2}\left( \frac{\alpha _sN_c}\pi \right) ^3
 \left( \frac{395}{27}-2\zeta (3)-
 \frac{11}3\frac{\pi^2}6+\frac{n_f}{N_c^3}\left(\frac{71}{27}-\frac{\pi^2}
 9\right)\right).
\label{gamma expansion}
\end{eqnarray}
The result for the three-loop correction was confirmed later in \cite{Vogt:2004mw},
\cite{Kotikov:2004er}.

\subsection{BFKL in N=4 SUSY}
Impressive and surprising results connected with BFKL  were obtained by Lev in maximally extended supersymmetric Yang-Mills theory (N=4 SYM). 
The eigenvalues of the NLO BFKL kernel were calculated in \cite{Kotikov:2000pm}. 
They  turned out to be much simpler than in QCD: instead of (\ref{chi1gamman}) 
the one-loop correction takes the form  
\be
4\chi^{(1)}\left(\gamma, n\right)  \ = -2\Phi\left(  n,\gamma\right) -2\Phi\left(  n,1-\gamma\right) +6\zeta(3) +\psi''(\gamma +\frac{n}{2})+\psi''(\gamma +\frac{n}{2})~. \label{chisusy}
\ee
A remarkable fact is the disappearance  of non analytical in conformal spin  $n$ terms.
Moreover,  all functions entering in (\ref{chisusy})  have the property of
maximal transcendentality~\cite{Kotikov:2004er}. The maximal transcendentality of an expression means by definition, that the special functions and numbers with lower complexities do not contribute to it.  By definition  $\psi(x)$ has the transcendentality equal to 1, the transcendentalities
of $\psi^{(n)}$ and $\zeta (n+1)$ are $n+1$ and the additional poles in the sum over $k$
increase the transcendentality of the function $\Phi(M)$ up to 3. Lev put forward a remarkable hypothesis about growing with order of perturbation theory   maximal transcendentality of the eigenvalues of the BFKL kernel and   
the anomalous dimensions of twist-2 operators in $N=4$ SUSY. This hypothesis is not disproved and widely used now. Another of Lev's remarkable hypothesis is the hypothesis 
that in $N=4$ SUSY the Pomeron coincides with the Reggeized graviton, which
gives a possibility to calculate its intercept at large coupling constants. Maximal
transcendentality together with integrability allow one to find the anomalous dimensions of twist-2 operators in this model up to 4 loops. 

\section{Future development}
Evidently, the next step in the development of the BFKL approach should be the next-to-next-to-leading logarithmic approximation (NNLLA).  Unfortunately, in QCD the  BFKL equation in the next-to-next-to-leading approximation is  not yet obtained,  
although in maximally extended supersymmetric Yang-Mills theory  with large number of colours (planar N=4 SYM) impressive  results have been  obtained \cite{Gromov:2015vua} - \cite{Alfimov:2018cms}. 

\subsection{NNLLA BFKL in planar N=4 SYM}
In \cite{Gromov:2015vua}   the next-to-next-to-leading order (NNLO) corrections to the BFKL Pomeron eigenvalues for conformal spin $n=0$  were obtained. The method of calculation  was quite different from the one  used in the BFKL approach,  in which, first of all,  the kernel of the BFKL equation is calculated using the unitarity and analyticity.  
It was based on the integrability of the planar N=4 SYM and  on the  observation of L. N. Lipatov and A.V. Kotikov \cite{Kotikov:2000pm} that BFKL eigenvalues and anomalous dimensions of twist-2 operators are connected in planar N=4 SYM by analytic continuation.  The problem of calculation of the anomalous dimensions 
was  solved by the  quantum spectral curve method (QSC) \cite{Gromov:2013pga}, \cite{Gromov:2014caa} giving  a finite set of Riemann-Hilbert equations for exact spectrum of planar N=4 SYM theory.  Applicability of this method to calculation of the LO BFKL Pomeron eigenvalues was demonstrated in \cite{Alfimov:2014bwa}. Another significant assumption, besides the applicability of the QSC, was the possibility 
to write the relation between the spin $S$ of the twist-two  operator  $Tr(ZD_+^S Z)$  and  its full conformal dimension $\Delta = 2+S+\gamma$, $\gamma$  is the anomalous dimension, in the form 
\be
j(\Delta) = 2+S(\Delta)=1+\sum\limits_{n=1}^\infty (\frac{g^2N_c}{(4\pi)^2})^{n}\left[F_n\left(\frac{\Delta-1}{2}\right)+F_n\left(\frac{-\Delta-1}{2}\right)\right]~, 
\ee
where $F_n(x)$ are simple linear combinations of the nested harmonic sums with the same transcendentality  $n$
\be
S_{a_1,a_2,\dots,a_n}(x)=\sum\limits_{y=1}^x\frac{(\textrm{sign}\left(a_1\right))^y}{y^{|a_1|}}S_{a_2,\dots,a_n}(y)\;\;,\;\;S(x)=1\;. \label{harmonic sums}
\ee
The possibility of such representation  for the LO and NLO eigenvalues with 
\be
F_1=-4 S_1, 
\frac{F_2}{4}=-\frac{3}{2}\zeta_3+\pi^2\ln 2+\frac{\pi^2}{3}S_1+2S_3+\pi^2 S_{-1}-4 S_{-2,1}
\ee
was shown in  \cite{Costa:2012cb}. 
The three-loop result obtained in \cite{Gromov:2015vua} is: 
\[
\frac{F_3(x)}{256}=-\frac{5
   S_{-5}}{8}-\frac{S_{-4,1}}{2}+\frac{ S_1
   S_{-3,1}}{2}+\frac{S_{-3,2}}{2}-\frac{5S_2
   S_{-2,1}}{4}   +\frac{S_{-4} S_1}{4}+\frac{S_{-3}S_2}{8} 
\]  
\[ 
+\frac{3
   S_{3,-2}}{4}-\frac{3S_{-3,1,1}}{2} -S_1
   S_{-2,1,1}   +S_{2,-2,1}+3 S_{-2,1,1,1}-\frac{3S_{-2}
   S_3}{4} -\frac{S_5}{8}
\]
\[    
+\frac{ S_{-2} S_1 S_2}{4}   +\pi ^2\left[\frac{S_{-2,1}}{8}  -\frac{7S_{-3}}{48}
   -\frac{S_{-2}S_1}{12}
   +\frac{S_1 S_2}{48}  \right]\!-\!\pi ^4 \left[\frac{2S_{-1}}{45} -\frac{  S_1}{96}\right]
\]
\[ 
   +\zeta _3\left[-\frac{7S_{-1,1}}{4}
   +\frac{7S_{-2}}{8}
   +\frac{7S_{-1}
   S_1}{4}  -\frac{S_2}{16}\right]
   \]
   \[ 
   +\left[2
   \text{Li}_4\!\!\left(\tfrac{1}{2}\right)
   -\frac{ \pi
   ^2  \log ^2\!2}{12}+\frac{\log
   ^4\!2}{12}
\right]
   \left(S_{-1}- S_1\right)
   \]
\be 
   +\frac{\log ^5\!2}{60}-\frac{\pi ^2
   \log ^3\!2}{36} -\frac{2 \pi ^4 \log 2}{45}
   -\frac{\pi ^2
   \zeta _3}{24}+\frac{49 \zeta _5}{32}-2
   \text{Li}_5\!\!\left(\tfrac{1}{2}\right)\;. \label{Gromov}
\ee
The eigenvalues $\omega(\gamma) = j(i\nu)-1$  at $\gamma = \frac12+i\nu$ correspond to complex values of $\Delta$ and require analytical continuation  of the harmonic sums. Corresponding  prescriptions can be found in 
\cite{Kotikov:2005gr} -\cite{Blumlein:2009ta}.

The approach used in \cite{Velizhanin:2015xsa}  was also  different from the BFKL approach. The  NNLO corrections to the Pomeron eigenvalues for conformal spin $n=0$  were obtained from the constraints, coming from the six-loop anomalous dimension of twist-2 operators and  large-gamma limit. The results agree with \cite{Gromov:2015vua}. 

The calculations of  the three-loop corrections in   \cite{Caron-Huot:2016tzz} is closer to the BFKL approach since the corrections are calculated firstly to the BFKL kernel (in the Pomeron channel). But the method of the calculation is quite different from the dispersive method used in the BFKL approach. Moreover, the kernel is calculated not in the momentum representation, but in the impact parameter space. 
The method used is based on  established in \cite{Caron-Huot:2015bja} connection of the BFKL logarithms  with so-called non-global logarithms \cite{Dasgupta:2001sh}, \cite{Banfi:2002hw} the physics of soft wide-angle radiation. 

In \cite{Alfimov:2018cms} a general non-perturbative framework for the BFKL spectrum of planar N = 4 SYM was developed. As well as in \cite{Gromov:2015vua}, it is based on the QSC method, but the spectrum  was studied at arbitrary values of conformal spin n.

\subsection{NNLLA  in the BFKL approach}
The results described above were obtained in $N=4$ SUSY  model with unique properties and concern only the BFKL kernel. Recall that amplitudes are given by convolution of impact factors of scattering particles and the Green's function, which is determined by the kernel, so that to calculate the amplitudes, it is necessary to know not only the kernel, but also the impact factors. The BFKL approach  gives an algorithm for calculating both components. Besides this,  the amplitudes of both elastic and inelastic processes are calculated  in the original (dispersive) BFKL approach.  Therefore, its development in the NNLLA is very desirable. And finally,  it is highly desirable to develop this approach in QCD.  Unfortunately, this development  is facing great difficulties. 

Remind that original scheme of  derivation of the BFKL equation  looks as follows. Elastic scattering amplitudes   are calculated using   the  $s$-channel unitarity and analyticity. In the LLA and  in the NLLA,  only amplitudes  having the  pole Regge form contribute to unitarity relations. It permits to present  the $s$-channel discontinuities  of elastic processes  as the convolutions of   impact factors of scattering particles and  the   Green's function for two
interacting Reggeized  gluons. All these components     are expressed in terms of the Reggeon vertices and trajectory. 

If this scheme were applicable, for derivation of the BFKL equation in the NNLLA  would be sufficient  to calculate three-loop corrections to the trajectory, two-loop corrections to the vertex of one gluon production, one-loop corrections to the vertices  of two gluon  and quark-antiquark production and  to find in the Born approximation new vertices:  for  three gluon and quark-antiquark-gluon production.

However, this scheme is based on the pole Regge form of amplitudes which are used in the unitarity relations (\ref{elastic discontinuity}). It  can not be used  the NNLLA.  In this approximation two large logarithms can be lost in the product of two amplitudes in the unitarity relations (\ref{elastic discontinuity}).  It can be done  losing either one logarithm in each of the amplitudes or both logarithms in one of the amplitudes. 

In the first case it becomes necessary to take into account the imaginary parts of amplitudes with negative signature as well as amplitudes with positive signature, which both  are suppressed by one degree of the logarithm.  Note that the simple factorized  form (\ref{A 2-2+n}) is valid only for the real part of the amplitudes and must be corrected to account for the imaginary parts, which creates certain difficulties. But more difficulties are connected with  the second case.   In this  case  one of the amplitudes in (\ref{elastic discontinuity}) must  be  taken in the LLA  and the other in the NNLLA. Since the amplitudes in the LLA are real, only real parts of the NNLLA amplitudes are important in this case. But even for these parts  the pole Regge form becomes inapplicable because of the contributions of the three-Reggeon cuts  which  appear in this approximation.  

The first observation of the violation the pole Regge form was done  \cite{DelDuca:2001gu} in the  high-energy limit of the  results of direct two-loop calculations of the two-loop amplitudes for  $gg, gq$ and $qq$  scattering.  
Then,  the terms breaking the pole Regge form in two and three-loop  amplitudes of elastic scattering  were found in \cite{DelDuca:2013ara, DelDuca:2013dsa, DelDuca:2014cya} using the  techniques of infrared factorization.

It is necessary to say that, in general, breaking of the pole Regge form is not a surprise.  It is well known that  Regge poles in the complex angular  momenta plane generate Regge cuts. Moreover, in amplitudes with positive signature the  Regge cuts appear already in the LLA. In particular, the BFKL  Pomeron  is the two-Reggeon cut in  the complex angular  momentum plane.  But in amplitudes with  negative signature Regge cuts appear only in the NNLLA.  It is  natural to expect that the observed violation of the  pole Regge form can be  explained by their contributions.  

Indeed, all known cases of breaking of the pole Regge form are now  explained by the three-reggeon cuts \cite{Fadin:2016wso, Caron-Huot:2017fxr}. Unfortunately, the approaches and  the explanations used in these papers are different. Their  results 
coincide   in three loops  but may diverge in more loops. It requires further investigation.   

Consideration of three-reggeon cuts in many-particle amplitudes  is an even more complicated problem.

\section{Conclusion}
L.N. Lipatov stood at the origins of the BFKL approach and played a prominent role in its development. He left a huge legacy to his disciples and followers. Unfortunately, now they are deprived of his highest skill, and the depth and originality of his ideas.

\end{document}